\documentclass[10pt,aps,pra,twocolumn,superscriptaddress]{revtex4-2}
\usepackage[utf8]{inputenc}
\usepackage{soul}
\usepackage{textcomp}
\usepackage[english]{babel}
\usepackage{graphicx}
\usepackage{dcolumn}
\usepackage{bm}
\usepackage{amsmath}
\usepackage{verbatim}     
\usepackage[dvipsnames]{xcolor}       
\usepackage[normalem]{ulem}   
\usepackage{bbold}
\usepackage{mathrsfs}
\usepackage[mathscr]{eucal}
\usepackage{hyperref}
\usepackage{float}
\usepackage{multirow}
\usepackage{array}

\usepackage{physics}
\usepackage{braket}
\usepackage{mathrsfs}

\hypersetup{
   pdfpagemode=UseNone,
   pdfstartpage=1,
   pdfmenubar=true,
   pdftoolbar=true,
   colorlinks = true,
   linkcolor=blue,
   citecolor=blue,
   urlcolor=blue,
   bookmarksopen=false}
\newcolumntype{Y}{>{\centering\arraybackslash}X}

\renewcommand{\abs}[1]{|#1|}
\newcommand{\qo}[1]{``#1''}

\renewcommand{\epsilon}{\varepsilon}

\def\VR{\kern-\arraycolsep\strut\vrule &\kern-\arraycolsep}
\def\vr{\kern-\arraycolsep & \kern-\arraycolsep}
\usepackage{sistyle}
\usepackage{soul}
\definecolor{lightblue}{RGB}{185,210,248}
\definecolor{cblue}{RGB}{24,49,121}
\definecolor{fgreen}{RGB}{81,192,122}
\definecolor{tyellow}{RGB}{229,228,90}
\sethlcolor{lightblue}

\begin{document}

\title{Investigating the Performance of Adaptive Optics on Different Bases of Spatial Modes in Turbulent Channels}

\author{Rojan Abolhassani}
\affiliation{Nexus for Quantum Technologies, University of Ottawa, Ottawa, K1N 6N5, ON, Canada}

\author{Lukas Scarfe}
\affiliation{Nexus for Quantum Technologies, University of Ottawa, Ottawa, K1N 6N5, ON, Canada}

\author{Francesco Di Colandrea}
\affiliation{Nexus for Quantum Technologies, University of Ottawa, Ottawa, K1N 6N5, ON, Canada}
\affiliation{Dipartimento di Fisica \qo{Ettore Pancini}, Universit\`{a} degli Studi di Napoli Federico II, Complesso Universitario di Monte Sant'Angelo, Via Cintia, 80126 Napoli, Italy}

\author{Alessio D'Errico}
\affiliation{Nexus for Quantum Technologies, University of Ottawa, Ottawa, K1N 6N5, ON, Canada}
\affiliation{National Research Council of Canada, 100 Sussex Drive, Ottawa, K1A 0R6, ON, Canada}

 \author{Khabat Heshami}
\affiliation{Nexus for Quantum Technologies, University of Ottawa, Ottawa, K1N 6N5, ON, Canada}
\affiliation{National Research Council of Canada, 100 Sussex Drive, Ottawa, K1A 0R6, ON, Canada}

\author{Ebrahim Karimi}
\email{ekarimi@uottawa.ca}
\affiliation{Nexus for Quantum Technologies, University of Ottawa, Ottawa, K1N 6N5, ON, Canada}
\affiliation{National Research Council of Canada, 100 Sussex Drive, Ottawa, K1A 0R6, ON, Canada}
\affiliation{Institute for Quantum Studies, Chapman University, Orange, California 92866, USA}

\begin{abstract}
Quantum key distribution (QKD) allows secure key exchange based on the principles of quantum mechanics, with higher-dimensional photonic states offering enhanced channel capacity and resilience to noise. Free-space QKD is crucial for global networks where fibres are impractical, but atmospheric turbulence introduces severe states’ distortions, particularly for spatial modes. Adaptive optics (AO) provides a pathway to correct these errors, though its effectiveness depends on the encoding basis. Here, we experimentally evaluate a high-speed AO system for orbital angular momentum (OAM) modes, mutually unbiased bases (MUB), and symmetric, informationally complete, positive operator-valued measures (SIC-POVM) up to dimension $d=8$ in a turbulent free-space channel. While OAM states are strongly distorted, their cylindrical symmetry makes them optimally corrected by AO, yielding error rates below QKD security thresholds. MUB and SIC-POVM exhibit greater intrinsic robustness to turbulence but are less precisely corrected, though their performance remains within protocol tolerances. These results establish AO as a key enabler of secure, high-dimensional QKD and highlight the role of basis choice in optimizing resilience and correction.
\end{abstract}

\maketitle

\section{Introduction} 
Quantum key distribution (QKD) is a method of sharing a secure key over a public channel, where the sender, Alice, aims to establish a shared secure key with the receiver, Bob. The two parties, along with an eavesdropper, Eve, with access to the public channel, are only limited by the laws of physics\\
QKD can be implemented using different protocols, all based on the same fundamental principles, where the security of the communication is ensured by the impossibility of cloning unknown quantum states with 100\% fidelity~\cite{wootters1982nocloning}. These protocols mostly differ in the set of non-orthogonal bases they use. Some of the most well-known protocols include the BB84~\cite{bennett1984quantum}, the MUB (tomographic) protocol~\cite{bruss1998sixstate}, or the Singapore protocol~\cite{englert2004efficient} -- see \cite{Bouchard2018} for a comprehensive survey of various QKD protocols employing spatial modes of light. By going to higher dimensions, having access to a larger alphabet, it is possible to encode more bits per sifted photons. This suggests that higher dimensions can increase the amount of secure information density per photon~\cite{cerf2002dlevel}. Moreover, higher-dimensional QKD protocols exhibit greater resilience to noise, thereby increasing the security threshold required to obtain a positive key in the presence of potential eavesdroppers~\cite{ecker2023noise}.\\
While polarization is limited to two-dimensional (qubit) information encoding, spatial modes of light can be harnessed to encode quantum information in high-dimensional spaces~\cite{molina2001management,mair2001entanglement,molina2004qutrits}. A widely used spatial mode encoding makes use of superpositions of modes carrying orbital angular momentum (OAM). We can theoretically have unbounded orthogonal OAM-carrying states, and thus, encode information in $d$-dimensions rather than only two dimensions, hence the name qu$d$its instead of qu$b$its. These OAM carrying beams have a helical phase structure $e^{i\ell \phi}$, where $\phi$ is the azimuthal angle in cylindrical coordinates and $\ell$ is the topological charge, an integer representing the number of $2\pi$ phase windings around the beam axis. In addition, proper superpositions of these modes that form a new set of orthogonal modes can also be used as new bases for information carriers in QKD protocols~\cite{mirhosseini2015highdim}.\\
QKD using spatial modes of light has been demonstrated in different channels such as optical fibres,~\cite{sit2018quantum,cozzolino2019oam, wang2021high}, underwater~\cite{sit2017highdimensional,sit2018quantum,bouchard2018quantum,hufnagel2019characterization}, or free space~\cite{gibson2004freespace}. Free-space QKD can be implemented in places where fibre channels are not accessible, and is essential in scenarios such as satellite communications. One of the main obstacles faced in free-space QKD is the atmospheric turbulence, which can completely distort the phase of the signal beam, reducing or even compromising the security of the channel. \\
When the turbulence effects are weak enough to introduce mostly phase distortion and negligible amplitude modulations, we can compensate for the environmental noise by using fast adaptive optics (AO) systems~\cite{scarfe2025fast}. Since adaptive optics systems have limited spatial resolution, their performance may vary depending on the selected sets of spatial modes. Therefore, not all modes behave the same under the same circumstances.\\

Here, we investigate the behaviour and performance of different high-dimensional superpositions of the OAM modes in the form of mutually unbiased bases (MUB)~\cite{durt2010mub} and symmetric informationally complete positive operator-valued measure (SIC-POVM)~\cite{englert2004efficient} in turbulent channels. Our goal is to identify the optimal modes which are less vulnerable to turbulence and more effectively corrected using AO. Our findings indicate that AO systems achieve the highest correction performance for pure OAM modes; however, these modes are also the most sensitive to turbulence. In contrast, certain MUB superpositions demonstrate greater resilience to turbulence, though their error rates remain somewhat higher than those of the OAM basis. Nevertheless, the application of AO enables recovery of channel security, balancing resilience and correctability across different mode sets.\\

\begin{figure*}[!ht]
\includegraphics[width=\linewidth]{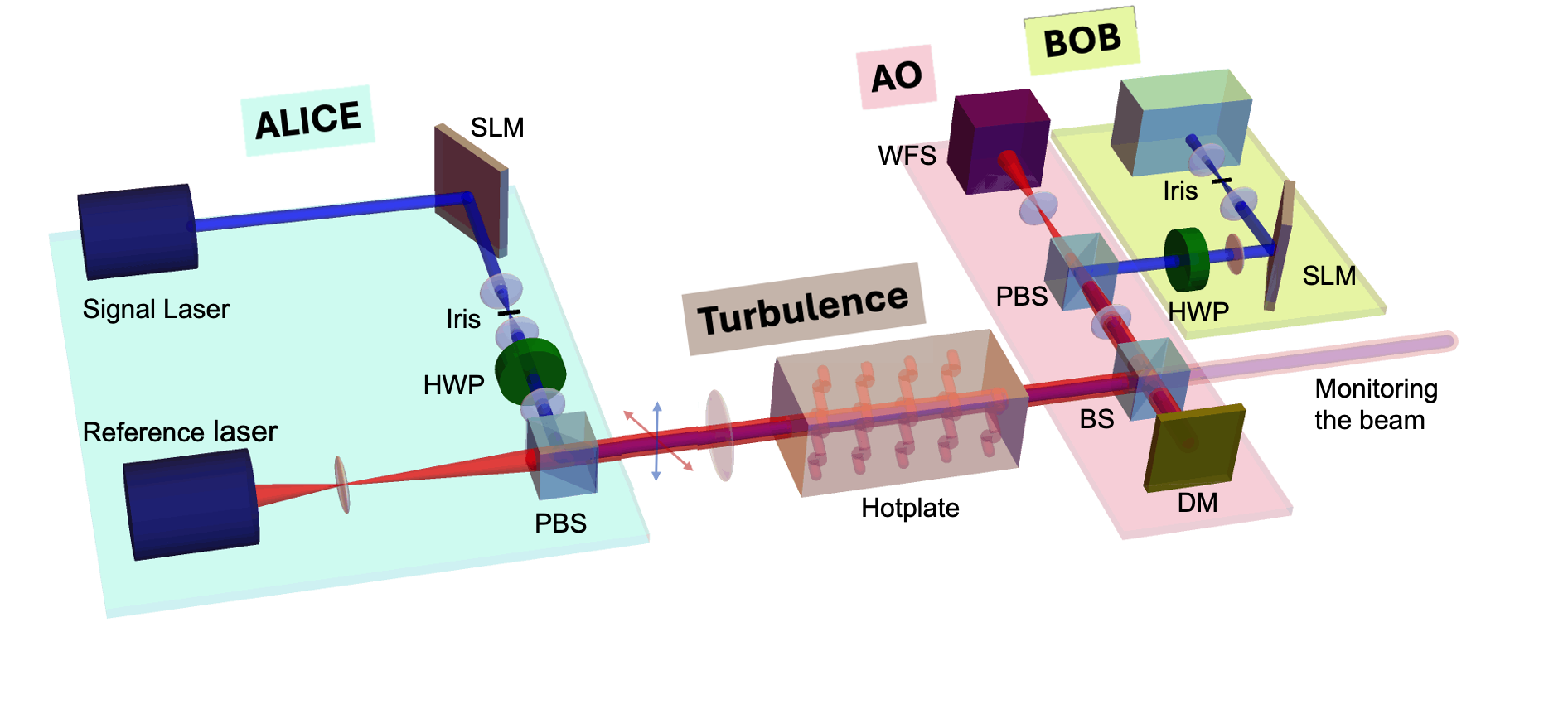}
\caption{Experimental setup for transmitting quantum states of light and testing MUB, SIC-POVM, and Angular states through turbulent media. Turbulence is generated using a controllable hotplate placed inside a 30-cm-wide glass tank. The signal beam, encoded with spatial mode information (MUB, SIC-POVM, or Angular) via a spatial light modulator (SLM), is combined with the reference beam at a polarizing beam splitter (PBS) and both co-propagate through the turbulence cell. The composite beam is then split by a 50:50 beam splitter (BS): one part is directed to a side wavefront sensor (WFS) to monitor the output wavefront, while the other part is sent to the adaptive optics (AO) system (from ALPAO). The AO system consists of a deformable mirror (DM) and a WFS operating in closed-loop feedback, where the DM corrects distortions measured by the WFS. After AO correction, the signal and reference beams are separated using a PBS. The signal component is then directed to a second SLM, which performs projective measurements of the spatial modes and couples the selected mode into a single-mode fibre (SMF) for detection, while the reference beam is sent to the WFS.}
\label{fig:setup}
\end{figure*}

\section{Results}
\subsection{Theory} 
We begin by outlining the definitions of the basis sets considered in this work. The set of modes which carry quantised non-zero orbital angular momentum, $\{\ket{\ell\}}_{\ell\neq 0}$, is considered as the \emph{logical} basis. The position $\mathbf{r}$ representation of these modes is considered to be Laguerre-Gaussian (LG) modes, $\braket{\mathbf{r}|\ell}:=\text{LG}_{\ell,p=0}(\mathbf{r})$, in transverse profile~\cite{allen1992orbital}. Here, $p$ is the radial index, but for simplicity, we consider it to be equal to zero; apart from a normalization factor, the expression for such a beam is given by $r^{\abs{\ell}}\,e^{i\ell\phi}\,e^{-(r/w_0)^2}$ with $w_0$ being the mode's beam waist.\newline

\noindent\textbf{Mutually Unbiased Bases:} Mutually Unbiased Bases (MUB) are a set of bases, each consisting of an orthonormal and complete set of elements, in which each vector basis or element has the following relation with other elements from the different basis sets:
\begin{equation}
   \abs{\langle{\beta_i^{(a)}}|{\beta_j^{(b)}}\rangle}^2 = \frac{1}{d}, \quad a\neq b
\end{equation}
where $\beta_i^{(a,b)}$ are elements of the basis $(a)$ or $(b)$, $i,j\in (1,d)$ and $d$ is the dimension of the Hilbert space. We investigate every MUB in dimensions $d={2,3,4,5,8}$, of which there are $d+1$ MUB. In each MUB, there are $d$ orthogonal vectors. Sets of $d+1$ MUB are known to exist in dimensions which are integer powers of a prime $d=p^n$~\cite{durt2010mub,WOOTTERS1989363}. Determining maximal sets of MUB in arbitrary dimensions remains an open problem~\cite{Raynal2011}; even in dimension six ($6 = 2 \times 3$), which is not a prime power, a complete construction is still unknown. In the simplest case where $d=2$, each MUB is given by the eigenvectors of the Pauli matrices. For polarization encoding, these (2+1=3) MUB correspond to horizontal and vertical polarization ($\{\ket{H},\ket{V}\}$), diagonal and anti-diagonal ($\{\ket{D},\ket{A}\}$), and left and right-circular polarizations ($\{\ket{L},\ket{R}\}$). In experiments concerning the OAM of light, we use the $\text{LG}_{\ell,0}$ modes as our logical bases and use them to build sets of MUB (considering $d$ is a power of a prime number). For dimension $d$, we use $\ell$ from $\{-d,-d+1,...,d-1,d\}$; however, $\ell =0$ was excluded, for even dimensions. The main reason was the potential non-negligible cross-talk with the natural mode of the single-mode fibre (SMF) during the final measurement. When using MUB, the outcome of measurement in a basis different from the logical basis has the same probability for each input OAM value; therefore, no information about its preparation could be obtained by the eavesdropper.\newline

\noindent\textbf{Angular Basis:} In the realm of high-dimensional QKD, the most commonly utilized secondary basis is generated by applying the discrete quantum Fourier transform to the logical OAM basis:
\begin{equation}
| \phi_{k} \rangle = \frac{1}{\sqrt{d}}\sum\limits_{j=0}^{d-1}e^{2\pi i \frac{jk}{d}}|j\rangle,
\end{equation}
where $|j\rangle = \frac{d}{2}+(l-1)\Theta(\ell)+\ell\Theta(-\ell)$  and $\Theta$ is the Heaviside function, i.e. we use $\ell$ from $\{-d,-d+1,...,d-1,d\}$ and exclude $\ell =0$ for even dimensions. This MUB is referred to as the angular mode set when the logical basis is given by OAM eigenstates and is characterized by basis elements with peak intensity located at a specific azimuthal angle $\phi=2\pi k/d$.
This angular basis yields the same states as the first MUB in the prime dimension cases. This Fourier-conjugate basis exhibits mutual unbiasedness with respect to the logical basis, indicating that a state prepared in one basis will produce uniformly random measurement outcomes when assessed in the other. Such complementary relationships are crucial for ensuring the security of QKD protocols, as they enable the detection of potential eavesdropping through heightened error rates in the conjugate measurements.\newline

\noindent\textbf{SIC-POVM:} Positive operator-valued measures (POVMs) are sets of positive semidefinite operators that describe generalized measurements on a quantum state. A POVM is said to be informationally complete (IC) if its measurement outcomes uniquely determine the state. Among these, a particularly important class is the symmetric informationally complete POVMs (SIC-POVMs), in which all elements share the same pairwise inner product, making them maximally efficient. Such sets provide a minimal spanning of the Bloch sphere and are widely regarded as optimal for state reconstruction~\cite{renes2004sicpovm}. It has been shown that in group-covariant cases, there are SIC-POVMs in all finite dimensions. SIC-POVMs in dimension $d$ are a set of $d^2$ normalized vectors $| \phi \rangle$ such that:
\begin{equation}
\abs{\langle \phi_i | \phi_j \rangle}^2 = \frac{1}{(1+d)}, \quad i \neq j\ .
\end{equation}
Completeness and informational completeness both follow from this property. One way of creating SIC-POVM elements is by applying the operator $D_{ij}$
\begin{equation}
D_{ij}=\omega^{jm}|k \oplus m\rangle\langle m|,
\end{equation}
to a Fiducial vector $| \phi \rangle$~\cite{Renes2004}. In this case $\oplus$ is the addition modulo $d$ and $\omega=\exp(2\pi i/d)$. In dimension $d=2$, the SIC-POVM corresponds to four pure states forming the vertices of a tetrahedron on the Bloch (Poincar\'e) sphere, which are mutually non-orthogonal. Such SIC-POVM bases have been employed in QKD protocols, for instance, in the Singapore protocol~\cite{englert2004efficient,Rambach2021}, which exhibits higher error tolerance compared with traditional MUB (tomographic)- and BB84-based schemes.
\begin{figure*}[t]
    \centering
	\includegraphics[width=\textwidth]{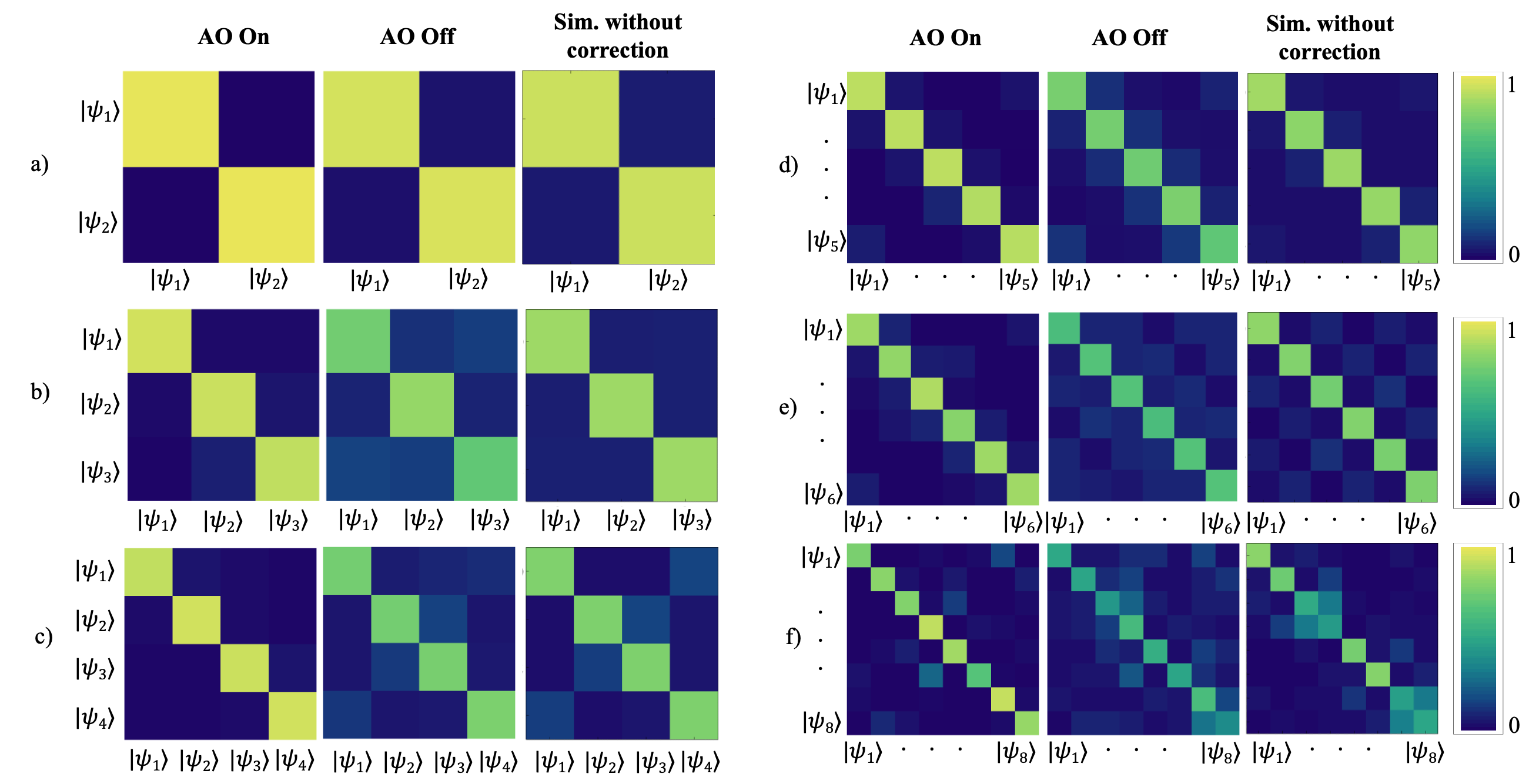}
    \caption{Cross-talk matrices of the first MUB for dimensions a) $d=2$, b) $d=3$, c) $d=4$, d) $d=5$, e) $d=6$, and f) $d=8$. Each panel displays three matrices for comparison: the first shows the experimental results obtained under turbulence with AO enabled (AO On), the second shows the corresponding experimental results under turbulence with AO disabled (AO Off), and the third presents numerical simulations of turbulence without correction (Sim). The comparison highlights the strong impact of turbulence on high-dimensional states and the significant improvement in mode fidelity achieved by AO correction, in close agreement with the simulated predictions.}
     \label{MUB:Cross}
\end{figure*}

\subsection{Experiment}
\noindent\textbf{Experimental setup:} The experimental setup consists of three main components: a sender (Alice), who prepares and encodes high-dimensional quantum information onto the spatial structure of photons; a turbulent channel, implemented using a controllable hotplate inside a 30 cm-wide glass tank; and a receiver (Bob), who ultimately decodes the transmitted states. The information-carrying beam is referred to as the signal beam. Alice prepares a continuous-wave 633 nm laser and directs it onto a phase-only spatial light modulator (SLM) to encode the desired state. By imprinting computer-generated holograms, the SLM modulates both the phase and the amplitude profile of the incoming beam~\cite{bolduc2013hologram,kirk1971phaseonly}, thereby creating high-dimensional spatial modes for transmission (see Fig.~\ref{fig:setup}).
In parallel, a second laser produces a Gaussian reference beam that is introduced into the system to probe and compensate for turbulence. This reference beam is prepared with a polarization state orthogonal to that of the signal, ensuring that the two beams can be deterministically separated at later stages of the experiment. The signal and reference beams are combined at a polarizing beam splitter (PBS) and propagate collinearly through the turbulence cell, where both experience identical wavefront distortions. The turbulence itself is generated by the convection currents of a controlled hotplate placed at the base of the glass tank, producing refractive-index fluctuations representative of atmospheric conditions.
After traversing the turbulence, the composite beam is directed to the AO system, consisting of a deformable mirror (DM, ALPAO) and a Shack–Hartmann wavefront sensor (WFS). A beam splitter (BS) divides the incoming light: one portion is directed to a second WFS for real-time wavefront measurements, while the DM reflects the other. WFS and DM are connected in a closed-loop control configuration, allowing the DM to actively reshape its surface in response to the measured aberrations and thereby compensate for turbulence-induced distortions. Following this correction, the reference and signal beams are separated by a second PBS: the reference beam is sent to the WFS for continuous monitoring, while the signal beam is directed to Bob.
At the receiver, Bob performs projective measurements on the signal beam over the chosen set of spatial modes. Here, the measurements are implemented using an intensity-flattening approach: a second SLM is programmed with phase masks that convert the incoming spatial modes into Gaussian-like profiles, which are then coupled into a SMF and detected~\cite{bouchard2018modes}. This method provides high mode selectivity and allows faithful discrimination between the high-dimensional states transmitted through the channel.
To characterize the level of turbulence introduced in the experiment, we compared the measured cross-talk matrices with theoretical predictions under different turbulence strengths. Agreement between simulation and experiment allowed us to estimate the effective refractive-index structure constant, $C_n^2 \approx 10^{-14.7}$~\cite{tatarski1961wave}, which is typical of moderate atmospheric turbulence. Further details of the turbulence modelling and numerical simulations are provided in the Supplementary Material.\newline

\noindent\textbf{Cross-Talk Matrix and QDER Measurements:} For each incoming basis, $\{| \psi_j \rangle\}$, Bob performs projective measurements of the signal beam onto the different modes, $\{|\langle \psi_i |\}$. In the case of MUB and angular (ANG) modes, the cross-talk is quantified by evaluating the projection probabilities $|\langle \psi_i | \psi_j \rangle|^2$ across all modes within the basis. To normalize the projective measurement matrix, each row is scaled so that its sum equals one. The resulting normalized matrix defines the cross-talk distribution:
\begin{equation}
C_{i,j}=\frac{|\langle\psi_i|\psi_j\rangle|^2}{\Sigma_{i=0}^{d-1}|\langle\psi_i|\psi_j\rangle|^2}.
\end{equation}

The key metric used to assess the communication performance and analyze the security is the quantum dit error rate (QDER). It is obtained by calculating the average of the diagonal elements of the normalized cross-talk matrix, and subtracting this value from the theoretical maximum of unity,
\begin{equation}
    \text{QDER} = 1-\frac{\Sigma_{i=0}^{d-1}C_{ii}}{d}.
\end{equation}
For the SIC-POVM, the cross-talk matrix is normalized so that each row sums to $d$, i.e.,
\begin{equation}
    C_{i,j}=d\times\frac{|\langle\psi_i|\psi_j\rangle|^2}{\Sigma_{i=0}^{d^2-1}|\langle\psi_i|\psi_j\rangle|^2}.
\end{equation}
In analogy to the MUB-based protocols, the QDER for SIC-POVMs can be estimated from the following expression:
\begin{equation}
    \text{QDER}=1-\frac{\text{Tr}[C_{ii}]}{d^2}.
\end{equation}
Here, $\text{Tr}(\cdot)$ denotes the trace operation, applied to the cross-talk matrix. The quantum error is then obtained by subtracting the average of the diagonal elements of the measured matrix from the corresponding theoretical value.
\newline

\noindent\textbf{Turbulence Compensation:} Phase distortions induced by turbulence are corrected using a closed-loop AO system from ALPAO. The AO system comprises three primary components: a Shack--Hartmann wavefront sensor (WFS), a control computer, and a deformable mirror (DM). Unlike earlier AO systems based on segmented mirror arrays, the DM employed here consists of a continuous reflective surface actuated by 97 electromagnetic elements. This design eliminates aberrations that would otherwise arise from inter-element gaps~\cite{tyson2015adaptive}. By locally adjusting the mirror surface, the DM reshapes the reflected beam in real time.  
The signal and reference beams are separated at a polarizing beam splitter (PBS), with the reference beam directed to the WFS. The Shack--Hartmann WFS samples the wavefront by measuring the displacement of focal spots relative to their ideal positions on the detector array. These displacements are then decomposed into Zernike polynomials, a widely used set of orthonormal functions for describing optical aberrations. Each polynomial corresponds to a distinct type of aberration (e.g., defocus, astigmatism, coma), enabling a compact representation of the measured wavefront distortion.  
This wavefront information is processed by the control computer, which generates the corresponding corrective commands for the DM. By applying the conjugate phase to the incoming distorted beam, the DM compensates for the turbulence-induced aberrations. The feedback loop operates at kilohertz rates: the WFS used here functions at 1~kHz (with a maximum capability of 5~kHz), while the turbulence dynamics are on the order of 100~Hz. Thus, the AO corrections are effectively applied in real time relative to the evolution of the turbulent channel.
\newline

\begin{figure*}[t]
    \centering
	\includegraphics[width=\textwidth]{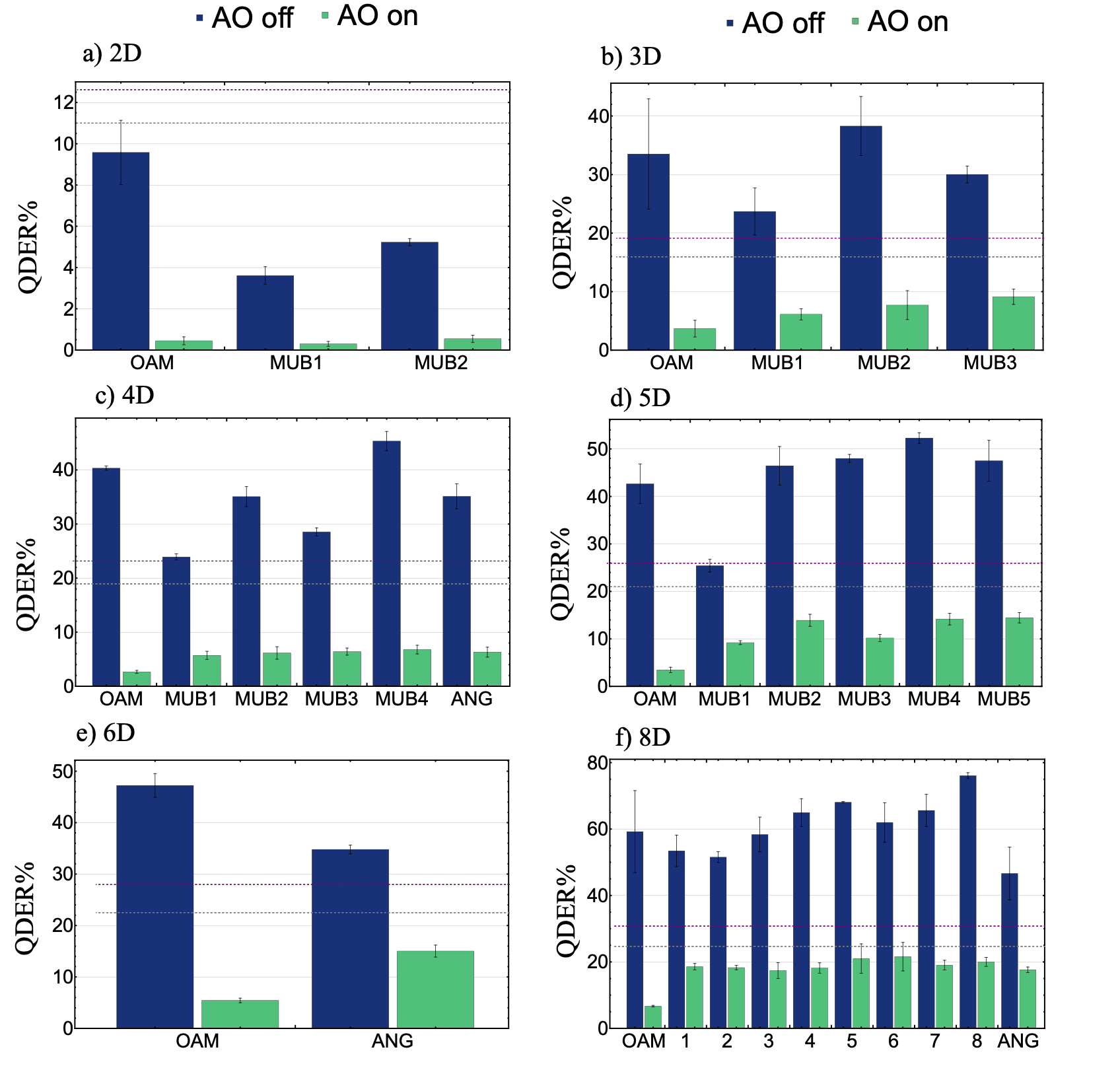}
	\caption{Quantum dit error rate (QDER) measured for different basis sets and Hilbert-space dimensions under turbulent conditions, with AO On (green) and with AO Off (blue). Panels a-f correspond to dimensions 2, 3, 4, 5, 6, and 8, respectively. Each bar shows the average QDER for OAM, MUB, and ANG bases (where applicable), with error bars representing statistical uncertainties. The dotted lines show the QDER thresholds for the BB84 (gray) and six-state (purple) protocols. The error lines on the bars indicate the standard deviation of the diagonal elements of each crosstalk matrix. The results demonstrate the strong effect of turbulence on high-dimensional states and the significant reduction in QDER achieved by AO correction.}
     \label{QDER:MUB} 
\end{figure*}

\noindent\textbf{Experimental results:} Using the setup described above, we performed experiments across different QKD protocols and Hilbert-space dimensions, evaluating the effects of turbulence and the corrective performance of the AO system. For each dimension, we measured the cross-talk matrices and extracted the corresponding QDER, both with AO disabled and enabled. Figure~\ref{MUB:Cross} shows the measured cross-talk matrices for different dimensions under both conditions, alongside numerical simulations based on the expected turbulence strength. From these results, QDER values were calculated for each basis and dimension. The corresponding results are summarized in Fig.~\ref{QDER:MUB}, which presents the QDER for MUBs, OAM modes, and ANG modes in dimensions $d=2,3,4,5,6,$ and $8$. We note that in the case of $d=6$, which is not a prime power, only three MUBs are known~\cite{PhysRevA.78.042312,PhysRevA.79.052316}; the analysis is therefore restricted to OAM and ANG modes, with the SIC-POVM case discussed separately.

Several clear trends emerge. First, higher-dimensional states are generally more susceptible to turbulence-induced errors. Second, activating the AO system significantly improves performance, reducing QDER values across all bases and dimensions. The AO correction is particularly effective for OAM modes, which, despite being highly distorted by turbulence, benefit strongly from the circular symmetry of the deformable mirror corrections. By contrast, certain MUB superpositions, while less corrected by AO, exhibit intrinsic robustness to turbulence due to their reduced spatial extent. These observations highlight an interplay between the geometric structure of the spatial modes and the corrective fidelity of the AO system. Modes with circularly symmetric profiles are more prone to turbulence but also more accurately corrected, whereas modes with localized lobes are less affected by turbulence but more challenging to correct. Intensity and phase profiles of these modes are shown in the Supplementary Material. 

The results further demonstrate that AO reduces the QDER most effectively in lower dimensions, bringing values well below the security thresholds for QKD protocols such as BB84 (two MUBs) and six-state (tomographic) schemes (three MUBs)~\cite{Bouchard2018}. Specifically, in dimension two (Fig.~\ref{QDER:MUB}(a)), the OAM and MUB states yield comparable QDERs; the ANG or first MUB basis emerges as the optimal choice in this regime. In dimension three (Fig.~\ref{QDER:MUB}(b)), which includes the Gaussian $\ell=0$ mode, we observe increased cross-talk, but the first MUB still performs better under turbulence and AO correction, making it a strong candidate alongside the OAM basis. In dimension four (Fig.~\ref{QDER:MUB}(c)), the Gaussian mode was excluded, the first MUB again shows superior resilience, while the fourth MUB is the most fragile, though still recoverable with AO. For dimension five (Fig.~\ref{QDER:MUB}(d)), the Gaussian mode component introduces additional noise, and although all bases are impacted, only the OAM and first MUB are well corrected. In dimension six (Fig.~\ref{QDER:MUB}(e)), where only three MUB sets are known, results are only reported for OAM and ANG modes. Finally, in dimension eight (Fig.~\ref{QDER:MUB}(f)), even with AO, some bases yield QDER approaching the security threshold, though OAM, ANG, and the first two MUBs remain viable candidates.  
Taken together, these results show that while turbulence strongly limits the fidelity of high-dimensional QKD, AO can substantially suppress the induced errors, bringing QDER values below security thresholds across a wide range of protocols and dimensions. This confirms the feasibility of secure, high-dimensional QKD over turbulent free-space channels when combined with an AO system. 

\begin{figure*}[!htb]
    \centering
	\includegraphics[width=\textwidth]{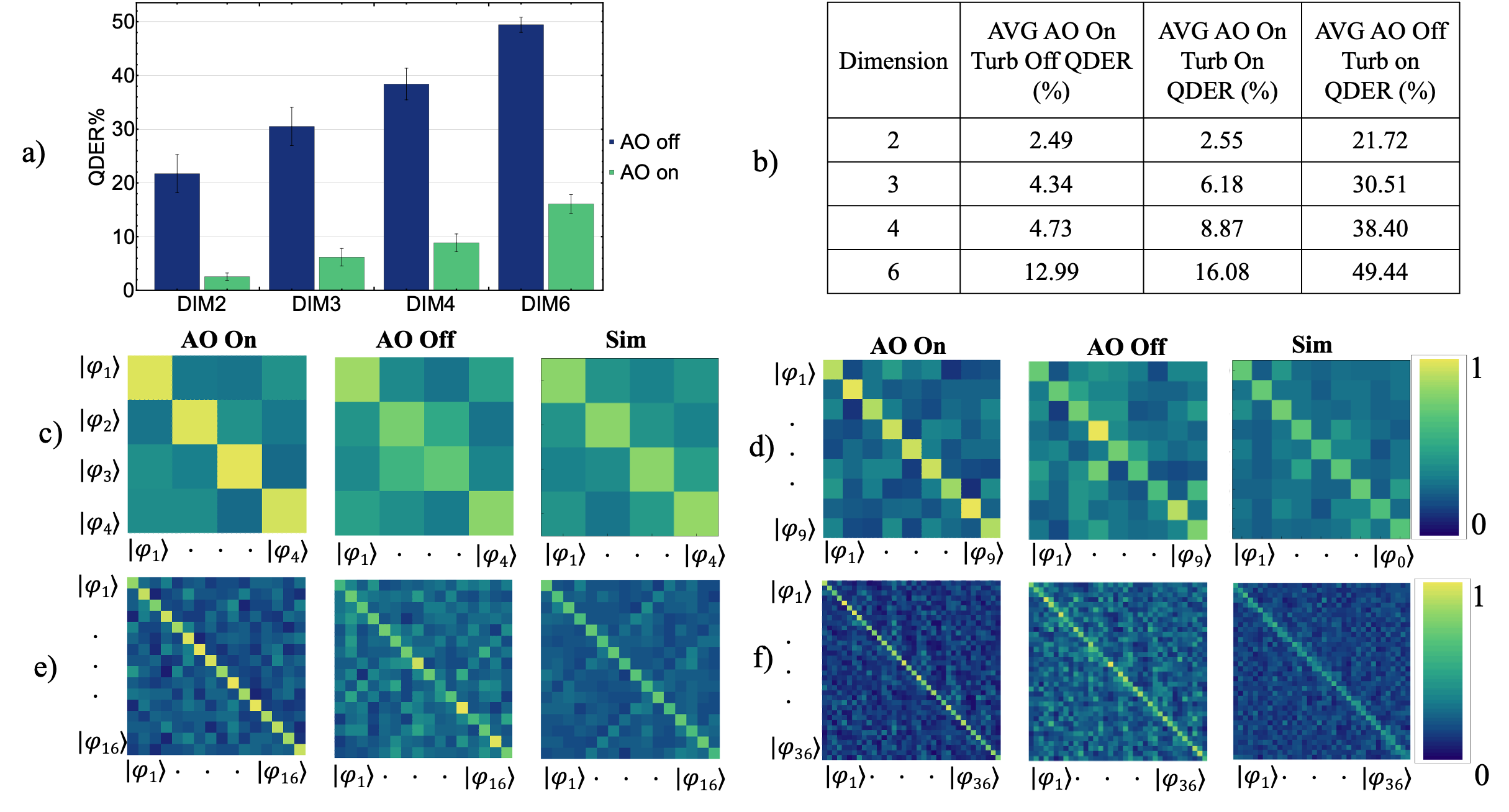}
     \caption{Cross-talk matrices and QDER for SIC-POVM modes under turbulence with and without AO correction. (a) Measured QDER for SIC-POVM modes transmitted through a turbulent channel, comparing AO-enabled (green) and AO-disabled (blue) scenarios. (b) Table summarizing the average QDER values obtained experimentally for SIC-POVM states in dimensions $d=2$, $d=3$, $d=4$, and $d=6$ under different conditions. (c–f) Experimental cross-talk matrices for SIC-POVM modes in dimensions 2, 3, 4, and 6, respectively, shown for AO enabled (AO On), AO disabled (AO Off), and numerical simulations of turbulence without correction (Sim). These results demonstrate the vulnerability of SIC-POVM states to turbulence and the significant reduction in error achieved by AO correction.}
     \label{SIC}
\end{figure*}

Beyond testing the maximal sets of MUB available in a given dimension $d$, it is also important to evaluate the capability of our AO system with other classes of states, thereby demonstrating its versatility for a broader range of quantum communication protocols. In particular, since no complete set of MUBs exists in dimension six, we consider alternative measurement frameworks such as SIC-POVMs, which in bi-dimensional space form the basis of the Singapore protocol~\cite{englert2004efficient}.
Our experimental results (Fig.~\ref{SIC}) show that SIC-POVM states are strongly affected by turbulence, leading to substantial cross-talk and elevated QDER. However, when AO correction is applied, the QDER is significantly reduced across all tested dimensions ($d=2,3,4,$ and $6$), with values falling below the theoretical security threshold of the Singapore protocol. This demonstrates that while SIC-POVMs are inherently fragile in turbulent environments, AO correction restores their viability for secure quantum key distribution. The cross-talk matrices further confirm this conclusion: with AO enabled, the measured mode overlaps approach the expected theoretical distributions, in close agreement with numerical simulations.
These results highlight both the susceptibility of SIC-POVM states to channel-induced distortions and the ability of AO systems to extend their practical utility in high-dimensional quantum communication protocols.

The security thresholds for different protocols are summarized in Table~\ref{all_table} for comparison~\cite{Bouchard2018}. Our results demonstrate that with AO correction, QKD can be implemented in turbulent channels while keeping error rates below the relevant security thresholds. At the same time, the results reveal that the AO system is not universal: it provides the most effective correction for specific mode families, particularly OAM modes. We also observe that certain modes exhibit greater intrinsic robustness to turbulence owing to their spatial localization.\\

\section{Conclusions}
In summary, we have systematically investigated the behaviour of all known MUBs and SIC-POVMs in multiple dimensions, up to $d=8$, under turbulence, and assessed the effectiveness of their correction using AO. These results provide practical guidance for selecting the most suitable basis sets under different channel conditions and for different QKD protocols. In particular, we find that in low-turbulence regimes, the OAM basis--although most strongly distorted by turbulence--benefits most from AO correction and thus emerges as the optimal channel for information transmission. This suggests that QKD schemes could be designed such that the logical (OAM) basis carries the majority of information, while MUB are employed primarily for security checks, thus leveraging an unbalanced usage of the bases. An example is provided by the recently proposed Fourier-qubit QKD~\cite{Scarfe2025HighDimensional}, where the OAM modes are applied more dominantly.
Beyond QKD, the implications of our results extend to other domains that rely on spatial modes of light. In biological imaging and microscopy, where turbulence and aberrations are often unavoidable, AO can mitigate distortions when using Fourier conjugates of OAM modes~\cite{Wu2025,Li2023,Pushkina2021}. Similarly, our findings indicate that AO correction can enhance the performance of vortex-beam-based coronagraphy~\cite{foo2005optical,palacios2006low,mawet2005annular}, further underscoring the broad applicability of these techniques.\newline

\noindent\textbf{Acknowledgment:} This manuscript has been proofread with the assistance of a large language model (LLM). The authors acknowledge support from the Canada Research Chairs (CRC) program, the National Research Council of Canada High-Throughput and Secure Networks (HTSN) Challenge Program, the Qeyssat User INvestigation Team (QUINT) Alliance Consortia, and the Alliance for Research and Applications of Quantum Network Entanglement (ARAQNE) Alliance Consortia Quantum grants. FDC acknowledges support from the PNRR MUR project PE0000023-NQSTI.\\

\noindent\textbf{Competing interests:} The authors declare no competing interests.\\

\noindent\textbf{Data Availability:} All data associated with this work are available upon request to the corresponding author.

\bibliographystyle{unsrt}
\bibliography{main}


\clearpage
\onecolumngrid
\renewcommand{\figurename}{\textbf{Figure}}
\setcounter{figure}{0} \renewcommand{\thefigure}{\textbf{S{\arabic{figure}}}}
\setcounter{table}{0} \renewcommand{\thetable}{S\arabic{table}}
\setcounter{section}{0} \renewcommand{\thesection}{S\arabic{section}}
\setcounter{equation}{0} \renewcommand{\theequation}{S\arabic{equation}}
\onecolumngrid

\begin{center}
{\Large Supplementary Material for: \\ Investigating the Performance of Adaptive Optics on Different Bases of Spatial Modes in Turbulent Channels}
\end{center}
\vspace{1 EM}

\subsection{Details on numerical simulations}
Our approach for numerical simulation of turbulence closely follows the method presented in Ref.~\cite{Jaouni:25}, which we briefly review here. In our simulations, aberrations are modeled as a single phase screen ${\xi(x,y)}$ located at the sender plane ${z=0}$: 
\begin{equation}
\tilde{E}(x,y,z=0)=E(x,y,z=0)e^{i\xi(x,y)},
\end{equation}
where $\tilde{E}$ and $E$ are the aberrated and input fields, respectively. The phase screen is generated as a superposition of Zernike modes~\cite{BornWolf}:

\begin{equation}
\xi(\rho,\phi)=\sum_{n,m}c_{n,m}Z_{n,m}(\rho,\phi),
\label{eqn:zernike}
\end{equation}
where $\rho$ is the normalized radial distance in the transverse plane: $\rho=\sqrt{x^2+y^2}/\text{Max}\left( \sqrt{x^2+y^2}\right)$, where the maximum is taken across the spatial window considered for the simulations. The Zernike mode of order $(n,m)$, with ${n\geq |m|>0}$, is defined as
\begin{equation}
Z_{n,m}(\rho,\phi)=R_{n,|m|}(\rho)\cdot\left(\Theta(m)\cos{m\phi}+\left(1-\Theta(-m)\right)\sin{m\phi} \right),
\end{equation}
where ${R_{n,|m|}}$ is the radial Zernike polynomial
\begin{equation}
R_{n,|m|}(\rho) = \sum_{k=0}^{\frac{n-|m|}{2}} \frac{(-1)^k (n-k)!}{k! \left( \frac{n+|m|}{2} - k \right)! \left( \frac{n-|m|}{2} - k \right)!} \rho^{n-2k}
\end{equation}
if $n-|m|$ is even and $0$ if $n-|m|$ is odd.
The weights of individual Zernike modes in Eq.~\eqref{eqn:zernike} are extracted from a zero-mean normal distribution depending on the turbulence conditions. Indeed, the variance ${\sigma^2_{n,m}}$ of the coefficient ${c_{n,m}}$ depends on the turbulence strength via the relation~\cite{Noll:76}
\begin{equation}
{\sigma^2_{n,m}}=\gamma_{n,m}\left(\frac{D}{r_0}\right)^{5/3},
\label{eqn:gamma}
\end{equation}
where $\gamma_{n,m}$ is a mode-dependent coefficient, $D$ is the receiver aperture, and $r_0$ is the Fried parameter. The latter can be determined from the $C_n^2$~\cite{tatarski1961wave}:

\begin{equation}
r_0=1.68\left(C_n^2Lk^2 \right)^{-3/5},
\end{equation}
where $L$ is the channel length and ${k=2\pi/\lambda}$ is the light wavevector. 

The propagation of the MUB states up to a distance $z$ without turbulence can be directly obtained from the analytical expressions of the OAM modes, which constitute the logical basis. When turbulence is present, however, there is no analytical expression for the propagation, and Fresnel diffraction of aberrated modes must be numerically simulated:
\begin{equation}
\tilde{E}(x,y,z)=\frac{e^{ikz}}{i\lambda z} e^{\frac{i\pi}{\lambda z}(x^2 + y^2)} \mathcal{F}(x,y,z),
\end{equation}
where
\begin{equation}
\mathcal{F}(x,y,z)=\iint \text{d}x' \text{d}y' \, \tilde{E}(x', y', 0) e^{i\frac{\pi}{\lambda z}(x'^2 + y'^2)}e^{\frac{2\pi i}{\lambda z}(xx'+yy')}.
\end{equation}

Finally, each element ${(i,j)}$ of the crosstalk matrices is obtained as the normalized mean of the overlap integrals between the ${i\text{-th}}$ and the ${j\text{-th}}$ modes, where the mean is performed over 100 random realizations of the chosen turbulence regime and the normalization is applied to each row independently. The overlap integral between a sent aberrated mode ${\tilde{E}_j}$ and a detected (non-aberrated) mode ${E_i}$ is given by
\begin{equation}
\mathcal{I}_{i,j}=\iint\text{d}x\text{d}y\,E_i^*(x,y,z)
\tilde{E}_j(x,y,z),
\end{equation}
where * denotes complex conjugation of the field.
Since optical propagation acts as a unitary transformation on the beam, the overlap integrals can be computed at any distance along the propagation. For convenience, the overlap is evaluated at $z=0$.

\goodbreak
\subsection{Table of Results}
\begin{table*}[ht]
\centering
\renewcommand{\arraystretch}{1}
\setlength{\tabcolsep}{12pt}
 
\renewcommand{\arraystretch}{1}
\begin{tabular}{|*{7}{>{\centering\arraybackslash}p{1.7cm}|}}

\hline
\textbf{Dimension}\rule{0pt}{1cm} & \textbf{Basis} & \textbf{AVG AO On Turb Off QDER(\%)} & \textbf{AVG AO On Turb On QDER(\%)} & \textbf{AVG AO Off Turb on QDER(\%)} & \textbf{BB84 Threshold
QDER(\%)} & \textbf{MUB protocol
Threshold QDER(\%)} \\
\hline

\multirow{3}{*}{\centering DIM~2} & MUB 0 (OAM) & 0.13± 0.03 &0.45 ± 0.19&9.58± 1.56 & \multirow{3}{*}{11.00} & \multirow{3}{*}{12.62} \\ \cline{2-5}
  & MUB 1 & 0.14± 0.09 & 0.30 ±0.12 & 3.61±0.43& & \\ \cline{2-5} &
 MUB 2 &0.32± 0.20 &0.55± 0.173 &5.23± 0.17 & & \\ \hline

\multirow{4}{*}{DIM~3} & MUB 0 (OAM) & 1.67± 0.51 & 3.68±1.42 & 33.51±9.41 & \multirow{4}{*}{15.95} & \multirow{4}{*}{19.14} \\ \cline{2-5}
 & MUB 1&5.11± 1.36 &6.13±0.96 &23.69±4.02 & & \\ \cline{2-5}
 & MUB 2 &6.93± 2.87 &7.70 ±2.470& 38.28±5.05& & \\ \cline{2-5}
 & MUB 3 &7.96± 1.58 & 9.11±1.30 &30.02±1.44 & & \\ \hline

\multirow{6}{*}{DIM~4} &MUB 0 
(OAM) & 1.10± 0.25&2.65 ± 0.28 &40.33± 0.38 &\multirow{6}{*}{18.93} & \multirow{6}{*}{23.17} \\ \cline{2-5}
 &MUB 1 &4.32± 0.92 &5.72± 0.74 &23.91± 0.56 & & \\ \cline{2-5}
 & MUB 2&5.22± 1.01 &6.16± 1.15 &35.07± 1.85 & & \\ \cline{2-5}
 &MUB 3 & 4.54± 0.40 & 6.41± 0.66& 28.54± 0.75& & \\ \cline{2-5}
 & MUB 4 &4.43± 0.99 & 6.78± 0.80& 45.34± 1.77& &\\ \cline{2-5}
 &ANG &4.35± 0.527 &6.31± 0.914 & 35.12± 2.31& & \\ \hline

\multirow{6}{*}{DIM~5} & MUB 0 (OAM) &1.074 ±0.21 &3.46±0.57&42.66±4.17& \multirow{6}{*}{20.99} & \multirow{6}{*}{25.94} \\ \cline{2-5}
 &MUB 1 &8.82±0.81 &9.21±0.39 &25.42±1.33 & & \\ \cline{2-5}
 &MUB2 &11.67 ±1.63 &13.92±1.26 &46.46 ±4.05& & \\ \cline{2-5}
 & MUB 3&8.73±0.48&10.19±0.78 &48.00±0.85 & & \\ \cline{2-5}
 & MUB 4&13.43±1.67 &14.16±1.22 &52.30±1.12 & & \\\cline{2-5}
 & MUB 5 & 14.01± 1.10 &14.46±1.07 & 47.52 ±4.28& & \\
 \hline

\multirow{2}{*}{DIM~6} &OAM &3.27± 0.20 &5.44± 0.448 &47.23± 2.31 & \multirow{2}{*}{22.50} & \multirow{2}{*}{27.97} \\ \cline{2-5}
 & ANG &14.65± 1.23 &15.03± 1.16 &34.79± 0.87 & & \\ \hline

\multirow{10}{*}{DIM~8} & MUB 0 (OAM)&2.68±1.14 &6.67±0.25 &46.64±7.94 & \multirow{10}{*}{24.70} & \multirow{10}{*}{30.77} \\ \cline{2-5}
 & MUB 1 &16.90±9.10 &18.57±0.95 &53.44±4.75 & & \\ \cline{2-5}
 & MUB 2 &16.40±5.56 &18.30±0.64&51.54±1.61& & \\ \cline{2-5}
 & MUB 3&17.81±1.95 &17.39±2.42&58.36 ±5.23& & \\ \cline{2-5}
 &MUB 4 &16.60±6.16 &18.17 ±1.56 &64.95 ±4.19& & \\ \cline{2-5}
 & MUB 5&22.06±3.58 &20.99 ±4.44&68.08 ±0.22 & & \\ \cline{2-5}
 & MUB 6& 22.54±2.14 &21.59 ±4.28 &61.99 ±5.95 & & \\ \cline{2-5}
 & MUB 7&17.96±1.67 &19.03 ±1.46 &65.60 ±4.84& & \\ \cline{2-5}
 &MUB 8 &18.56 ±1.72&20.00 ±1.34 &76.08 ±0.88 & & \\ \cline{2-5}
 &ANG &18.15±2.65 &17.64 ±0.82& 46.64 ±7.94 & & \\
 \hline

\end{tabular}

\caption{Table of experimental results showing Quantum dit error rates for all bases in different dimensions. The results refer to cases of adaptive optics turned on and off with turbulence on and off. The last two columns indicate the security threshold for doing QKD with different protocols like the BB84 and six-state or MUB protocol.}
\label{all_table}
\end{table*}

\clearpage

\subsection{Phase-Intensity Plots of MUB}
\begin{figure*}[!ht]
    \centering
	\includegraphics[width=\textwidth]{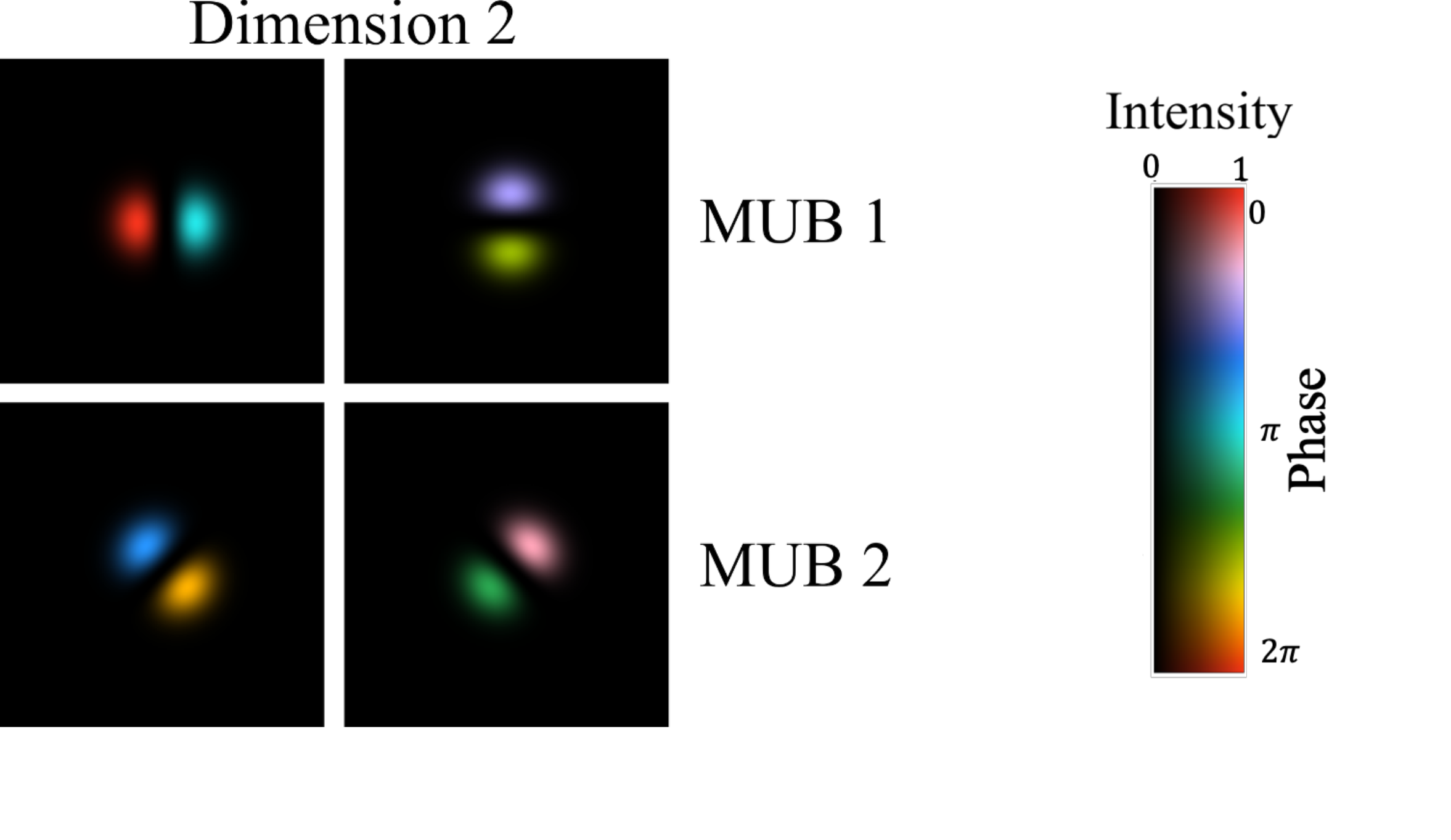}
	 \label{2D MUB}\caption{Phase and intensity plots of the MUB in dimension 2 using the OAM basis. Note that all phase-intensity plots use the same colour-scale as this plot.}
\end{figure*}

\begin{figure*}[!ht]
    \centering
	\includegraphics[width=\textwidth]{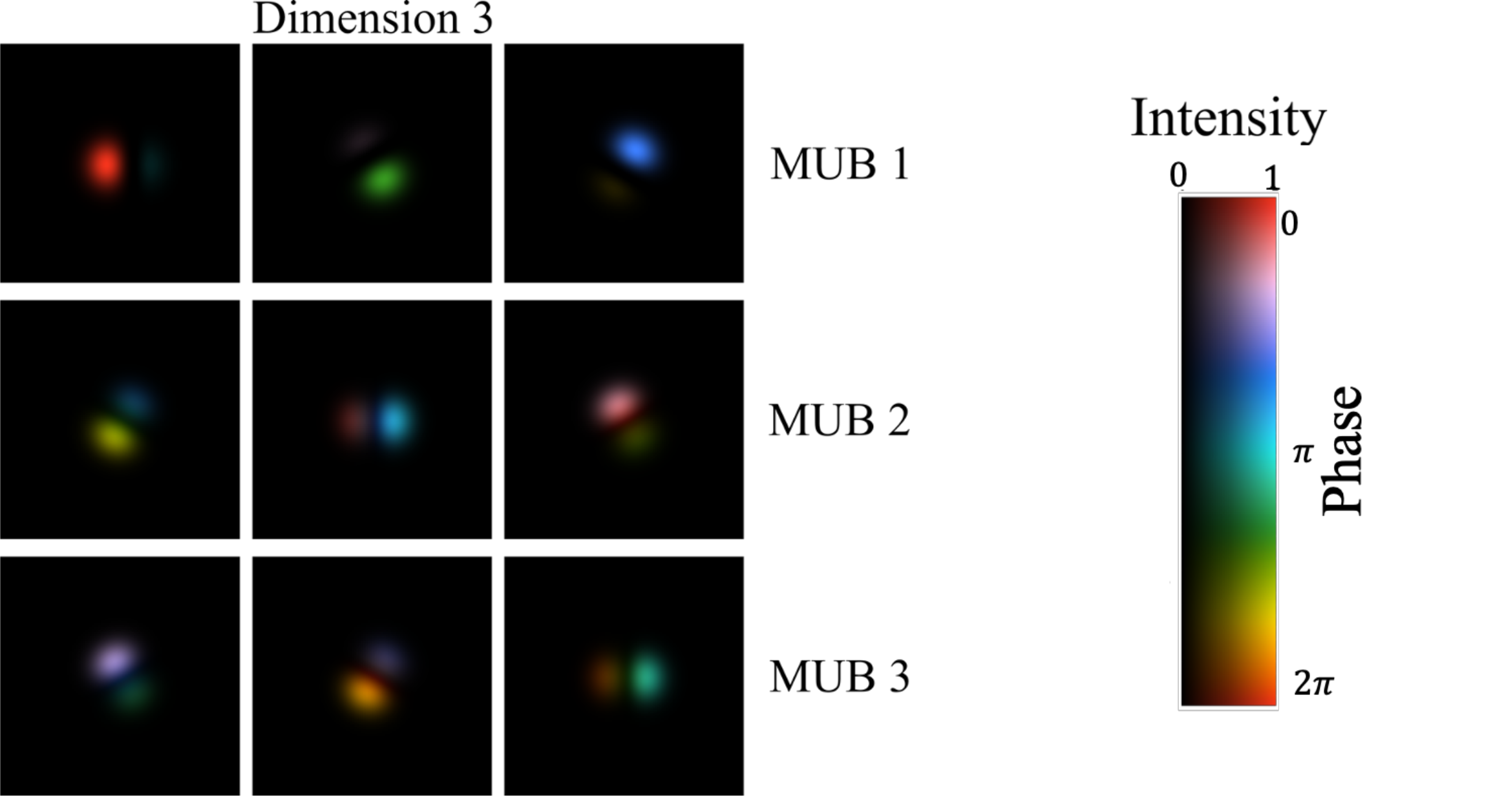}
	 \label{3D MUB}\caption{Phase and intensity plots of the MUB in dimension 3 using the OAM basis.}
\end{figure*}

\begin{figure*}[!ht]
    \centering
	\includegraphics[width=\textwidth]{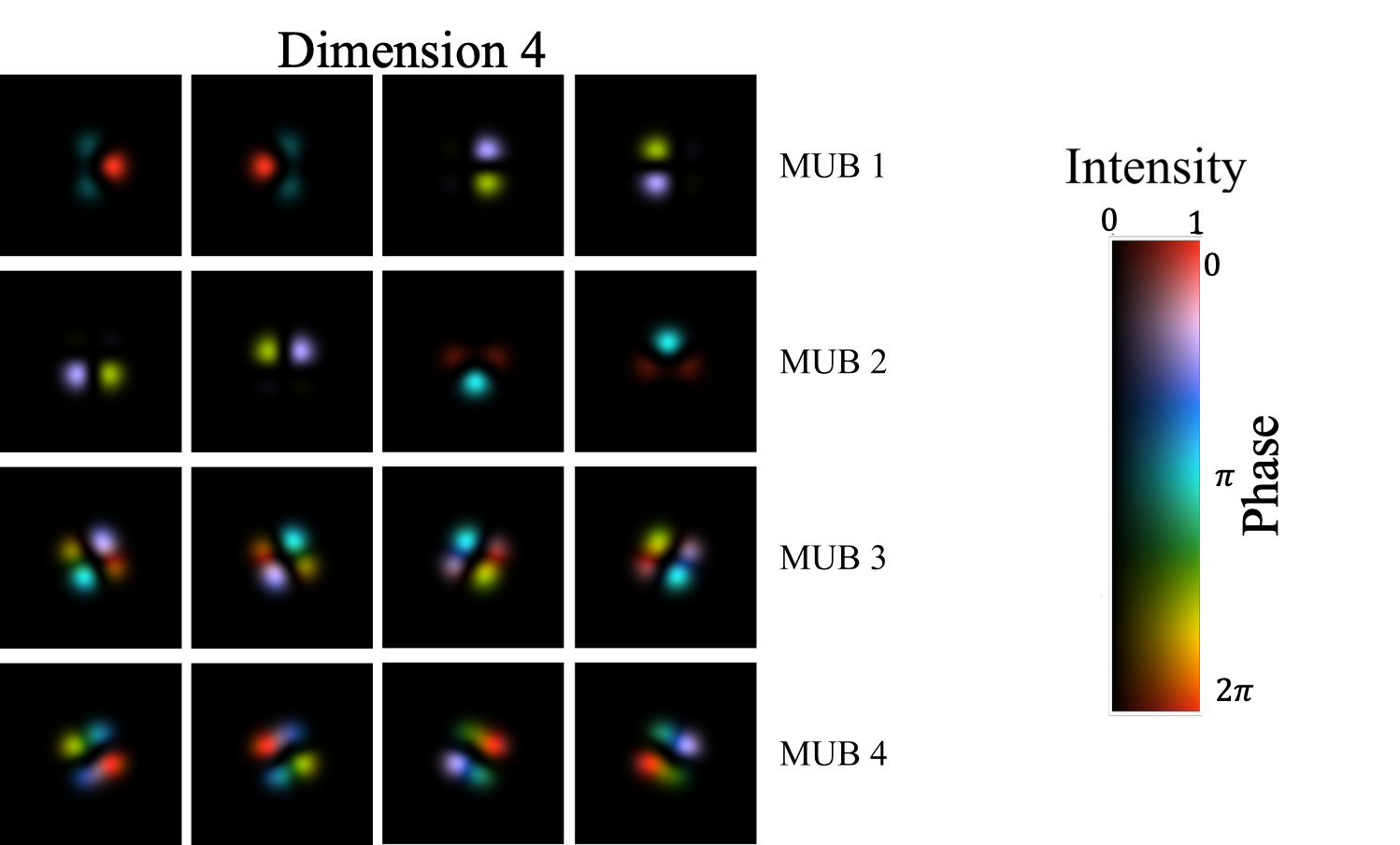}
	 \label{4D MUB}\caption{Phase and intensity plots of the MUB in dimension 4 using the OAM basis.}
\end{figure*}

\begin{figure*}[!ht]
    \centering
	\includegraphics[width=\textwidth]{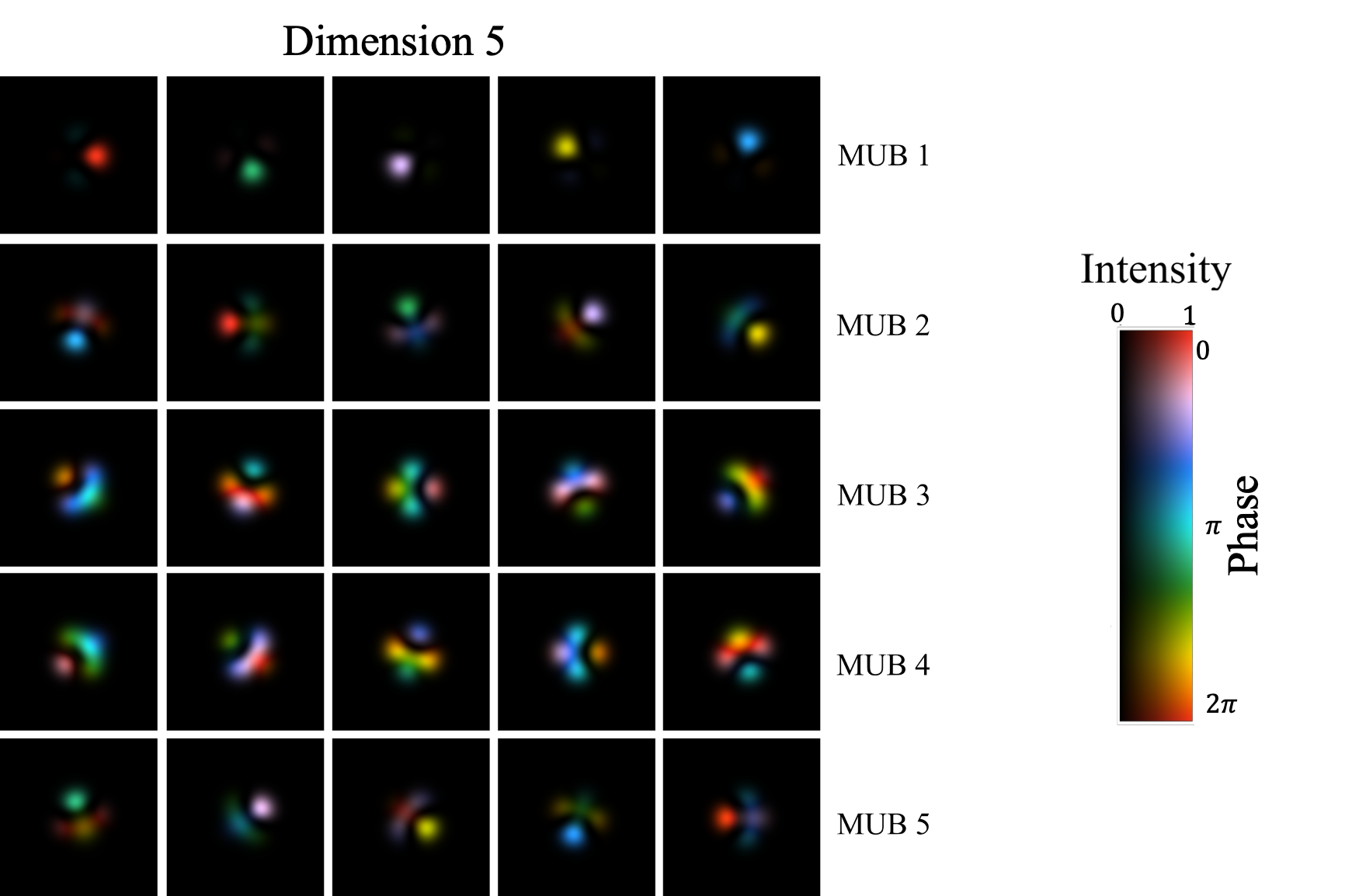}
	 \label{5D MUB}\caption{Phase and intensity plots of the MUB in dimension 5 using the OAM basis.}
\end{figure*}

\begin{figure*}[!ht]
    \centering
	\includegraphics[width=\textwidth]{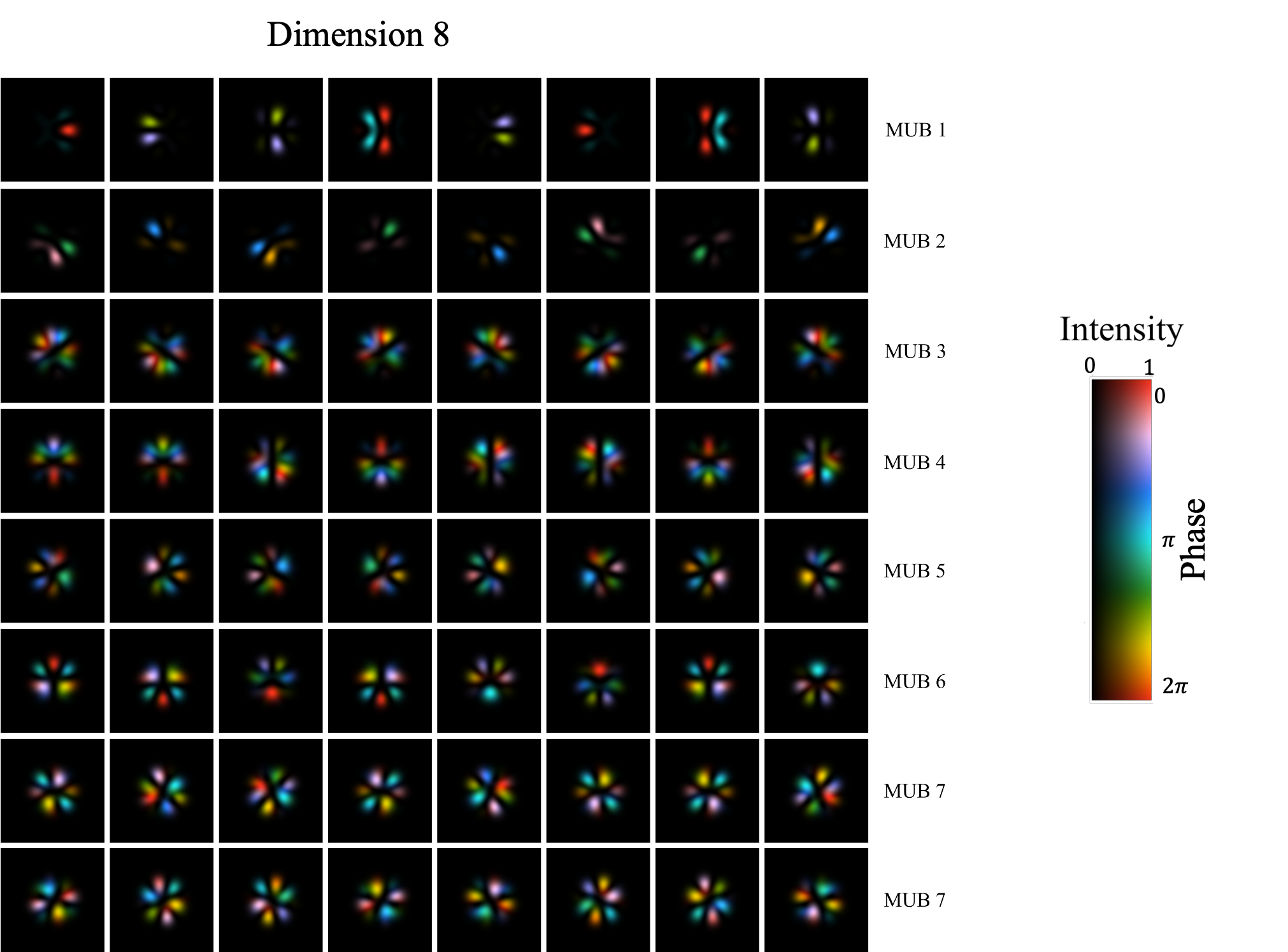}
	 \label{8D MUB}\caption{Phase and intensity plots of the MUB in dimension 8 using the OAM basis.}
\end{figure*}

\end{document}